\documentclass[conference,compsoc]{IEEEtran}

\usepackage{algorithm}
\usepackage{algorithmic}
\usepackage{graphicx}
\usepackage{booktabs}
\usepackage{listings}
\usepackage{xcolor}
\usepackage{subcaption}
\usepackage{hyphenat}
\usepackage{xspace}
\usepackage{amsmath}
\usepackage{siunitx}
\usepackage{enumerate}
\usepackage{placeins}
\usepackage{tikz}
\usepackage{url}
\usepackage{hyperref}
\hypersetup{
    colorlinks,
    linkcolor={black},
    citecolor={red!70!black},
    urlcolor={blue!70!black}
}
\usepackage{cleveref}
\usepackage[final]{microtype}

\usepackage[
	backend=biber,
	bibstyle=numeric,
	citestyle=numeric-comp,
	sortcites=true,
	url=false,
	urldate=iso,
	seconds=true,
	maxbibnames=1,
	minbibnames=1,
]{biblatex}
\DefineBibliographyStrings{english}{urlseen = {Accessed:}}
\newcommand\tool{\textsc{Walma}\xspace}

\newcommand{\cpp}{\mbox{C++}\xspace}

\newcommand\circled[1]{%
    \tikz[baseline=(char.base)]{
        \node[shape=circle,draw,inner sep=.5pt] (char) {#1};
}}

\makeatletter
\newcommand{\authorcell}[2]{\begin{tabular}{@{}c@{}}{\@IEEEauthorblockNstyle #1}\\
    \@IEEEauthorblockAstyle\begin{tabular}{@{}c@{}}#2\end{tabular}\end{tabular}}
\makeatother

\begin{document}
\title{\tool: Learning to See Memory Corruption in WebAssembly}

\author{\begin{tabular}{@{}c@{\hspace{2.5em}}c@{}}
\authorcell{Oussama Draissi}{\textit{paluno -- The Ruhr Institute for Software Technology}\\
University of Duisburg-Essen\\
Essen, Germany\\
oussama.draissi@uni-due.de} &
\authorcell{Mark G\"unzel}{\textit{paluno -- The Ruhr Institute for Software Technology}\\
University of Duisburg-Essen\\
Essen, Germany\\
mark.guenzel@stud.uni-due.de} \\
\multicolumn{2}{c}{\rule{0pt}{1.2em}} \\
\authorcell{Ahmad-Reza Sadeghi}{\textit{System Security Lab}\\
TU Darmstadt\\
Darmstadt, Germany\\
ahmad.sadeghi@trust.tu-darmstadt.de} &
\authorcell{Lucas Davi}{\textit{paluno -- The Ruhr Institute for Software Technology}\\
University of Duisburg-Essen\\
Essen, Germany\\
lucas.davi@uni-due.de}
\end{tabular}}

\maketitle

\begin{abstract}
The monolithic linear memory of WebAssembly (Wasm) creates a bidirectional attack surface: a memory corruption bug in a module can compromise the host, and a compromised host can corrupt the state of a trusted module.
Existing defenses either require custom runtimes or source code changes, and none can verify runtime integrity under an adversarial host.
We present \tool, the first attestation framework delivering \emph{resilience by reaction} against data-centric attacks that corrupt a region of memory. We render linear memory snapshots as grayscale images and classify them with a convolutional neural network.
The network reads the row-aligned structure that compiled code imposes on memory and detects corruption that byte and texture statistics miss.
Our main VGG-16 detector reaches \num{95}\,\%--\num{100}\,\% verdict accuracy on four of five CVE-affected applications without a single false alarm, and a \num{1.9}M-parameter ResNet, small enough to run inside the runtime on every host call, stays within a few points of it on the same four.
A Shannon-entropy analysis bounds which out-of-band edits the network can separate from valid state, and black-box evasion requires contiguous overwrites of hundreds of kilobytes to megabytes.
On the \num{53} WABench binaries, boundary attestation costs 1.07$\times$--1.69$\times$, scaling with a module's exposure to the host rather than with its runtime.
\tool thus makes linear-memory integrity a property a Wasm runtime can check continuously: end to end against compromised modules, and against untrusted hosts provided snapshots reach the verifier intact.


\end{abstract}


\section{Introduction}
\label{sec:intro}
WebAssembly (Wasm)~\cite{wasm} is a portable bytecode format that runs untrusted code inside a sandbox, has grown beyond its browser origins into a universal binary format for edge computing, serverless functions, smart contracts, and embedded systems~\cite{wasm_state_2025}.
Major platforms such as Figma, Google Earth, Zoom, Cloudflare Workers, and American Express's internal FaaS rely on it for performance-critical execution~\cite{figma_wasm,wasm_state_2025,amex_faas}.
Yet most Wasm modules are compiled from memory-unsafe languages such as C and \cpp~\cite{pitfalls,hilbig}.

A Wasm module holds all of its mutable state in one contiguous byte array with no internal memory protection~\cite{wasmSec}, its \emph{linear memory}.
A memory corruption bug is therefore rarely contained: an overflow in one buffer can silently corrupt any other structure in the module.
The sandbox isolates the module from its host, but the host implicitly \emph{trusts} data returned across the boundary.
Therefore, an attacker does not need to divert the control flow~\cite{rop1,rop2} to compromise the system: the corrupted \emph{data} it leaves behind is itself the exploit primitive, and this data-only escalation path~\cite{dop1,dop2} carries an in-module bug into a full host compromise~\cite{everything}.
A recent measurement study~\cite{wemby} finds thousands of websites that pass attacker-controllable Wasm data to unsafe sinks such as \verb|innerHTML|.
There, a corrupted Wasm-side value reaches the DOM as injected markup, turning an in-module memory bug into a working exploit.
The same monolithic memory model leaves the module defenseless in the opposite direction.
In cloud and edge deployments, a privileged host such as a malicious administrator or a compromised hypervisor~\cite{tee1,tee2,jattke2024zenhammer} can tamper with linear memory directly and subvert authorization checks before they reach the user.
Wasm thus faces a \emph{bidirectional} integrity problem that existing defenses leave open.
Isolation and hardening schemes~\cite{mswasm,rlbox,pkuwa,wave} demand custom runtimes, source changes, or specialized hardware; binary-rewriting memory-safety instrumentation~\cite{fuzzm,wbsan} is fragile under optimization and, running inside the module, powerless against an adversarial host.
ASLR~\cite{aslr} does not fit Wasm's flat memory model, and code- or control-flow attestation~\cite{swatt,pioneer,cflat,oat,litehax,atrium} misses the data-only attacks that drive real Wasm exploits~\cite{everything,chasms,wemby}.
What is missing is \emph{resilience by reaction}: a continuous, data-centric primitive that audits the module's mutable state itself, with a verifier placeable on either side of the trust boundary.

We present \tool, the first \underline{W}eb\underline{A}ssembly \underline{l}inear \underline{m}emory \underline{a}ttestation framework.
\tool treats the entire linear memory as a uniform attestation target and reduces integrity verification to a learned classification problem.
Snapshots of linear memory are rendered as fixed-width grayscale images, and a convolutional neural network~\cite{vgg} trained on fuzzer-generated benign and corrupted states classifies each snapshot at runtime.
The visual encoding exploits the row-aligned structure that compiled code imposes on memory---audio frames, pixel rows, allocator headers---and corruption therefore appears as a localized texture disruption a CNN detects without hand-engineered features.
The encoding is essential rather than cosmetic, beating the strongest non-CNN baseline by up to 22 F1 points (\Cref{sec:eval-detection}).
The attester--verifier interface is decoupled: the attester co-locates with execution, while the verifier runs inside a TEE, on the trusted browser host, or on an off-runtime accelerator.
Finally, because the benign training class alone defines what valid memory looks like, \tool's verdict does not depend on a sanitizer at inference.
On targets whose benign memory is structured, it also flags tampering that no program input triggers---direct memory writes by a malicious host, such as DMA~\cite{thunderclap}.
\tool thereby achieves four properties:
\circled{1}~\emph{data-centric attestation} that catches the memory footprint of data-only exploitation, which control-flow attestation misses;
\circled{2}~\emph{bidirectional applicability}, with a single attester--verifier protocol protecting a trusted module from a malicious host \emph{and} a trusted host from a compromised module;
\circled{3}~\emph{sanitizer independence at inference}, detecting tampering with no sanitizer in the loop; and
\circled{4}~\emph{tunable cost}, exposed through three instrumentation policies and three verifier backends, with a \num{1.9}M-parameter detector small enough to run inside the runtime on every host call.

Two scope decisions bound these properties.
First, \tool detects corruption that changes a region of memory, such as the heap and stack buffer overflows in our five targets.
A single flipped flag, length field, or byte falls below its detection floor and remains the domain of per-variable attestation~\cite{dialed,oneforall,dmanplay}.
Second, the verifier must receive snapshots intact.
In the browser scenario this holds by construction, because the trusted runtime captures memory itself, and our prototype realizes this path end to end.
In the cloud scenario the host is malicious, so the capture path must lie outside its control, for example through verifier-controlled DMA~\cite{dmanplay}.
Our prototype does not implement such a path, so here we evaluate the classifier assuming one.

On five CVE-affected applications, our main VGG-16 model~\cite{vgg} reaches \num{95}\,\%--\num{100}\,\% verdict accuracy on four of the five targets, without a single false alarm on benign inputs on any target.
A compact ResNet~\cite{resnet} is a few points less accurate but fast enough for high-frequency in-runtime attestation.
A black-box evasion study shows that flipping a verdict requires overwriting a large fraction of memory, hundreds of kilobytes to megabytes on our targets.
Across WABench's \num{53} binaries~\cite{wabench,polybench,mibench}, \tool exposes three operating points: boundary defense at every host interaction (1.07$\times$--1.69$\times$), which catches corruption before it escalates to an XSS or RCE;
per-function attestation (1.50$\times$--2.65$\times$);
and per-write attestation as a stress-test upper bound (1.78$\times$--7.57$\times$).
Counterintuitively, in-runtime CPU inference beats GPU offloading, because data transfer, not compute, dominates at this snapshot size.

In summary, we make the following contributions:
\begin{itemize}
    \item The design and implementation of \tool, the first linear-memory attestation framework for WebAssembly.
    \item A security evaluation on five CVE-affected applications including audio coding, raster conversion, bilevel image coding, and document parsing, with an input-level split that rules out memorizing inputs.
    \item A robustness characterization of the learned detector, rather than headline accuracy alone: an entropy-bounded detection floor that states which out-of-band edits, invisible to any sanitizer, the learned prior can and cannot separate from valid state, the black-box overwrite budget an adversary must spend to flip a verdict, and the accuracy cost of the \num{1.9}M-parameter model that makes in-runtime attestation affordable.
    \item A performance characterization over WABench's \num{53} binaries at three operating points, one per instrumentation policy: boundary defense at host calls, per-function, and per-write attestation as an upper bound.
\end{itemize}

We release \tool as open source at \url{https://github.com/uni-due-syssec/walma}.

\section{Background}
\label{sec:background}

\subsection{WebAssembly Security}
Wasm's security model~\cite{wasmSec} rests on software fault isolation (SFI)~\cite{sfi}, control-flow integrity (CFI)~\cite{cfi1}, and code immutability.
Together, they isolate the module from the host and rule out classical code-reuse~\cite{rop1} and code-injection~\cite{younan2012countermeasures} attacks.

\subsubsection*{The Monolithic Linear Memory}
Wasm's sandbox protects the host but not the module from itself.
Each module owns a single \emph{linear memory}: a contiguous, resizable byte array that co-locates stack, heap, and global data without runtime-enforced separation, and to which any instruction may freely read or write.
A buffer overflow in one region therefore corrupts adjacent structures in another, enabling the cross-region exploits that characterize Wasm memory bugs~\cite{everything,chasms,wemby}.
This one array is a single point of failure, but also a comprehensive attestation target.

\subsection{Remote Attestation}
Remote attestation lets a trusted \emph{verifier} judge the integrity of a remote, untrusted \emph{attester}.
Classical schemes verify static properties such as code identity~\cite{swatt,pioneer}; control-flow attestation (CFA)~\cite{cflat,oat,litehax,atrium} extends attestation to runtime behavior by cryptographically recording executed branches against the application's control-flow graph.
CFA still misses data-only attacks~\cite{sok_attestation,dop2}: decision variables in linear memory can be manipulated---by a memory corruption fault or by an external write---to subvert program logic without diverting a single branch.
This gap motivates \tool's complementary, \emph{data-centric} defense: validating the memory footprint itself rather than the execution path.

\section{Threat Model}
\label{sec:threat_model}

\begin{figure}[htbp]
    \centering
    \includegraphics[width=.7\columnwidth,trim={1.8cm 0 1.4cm 0},clip]{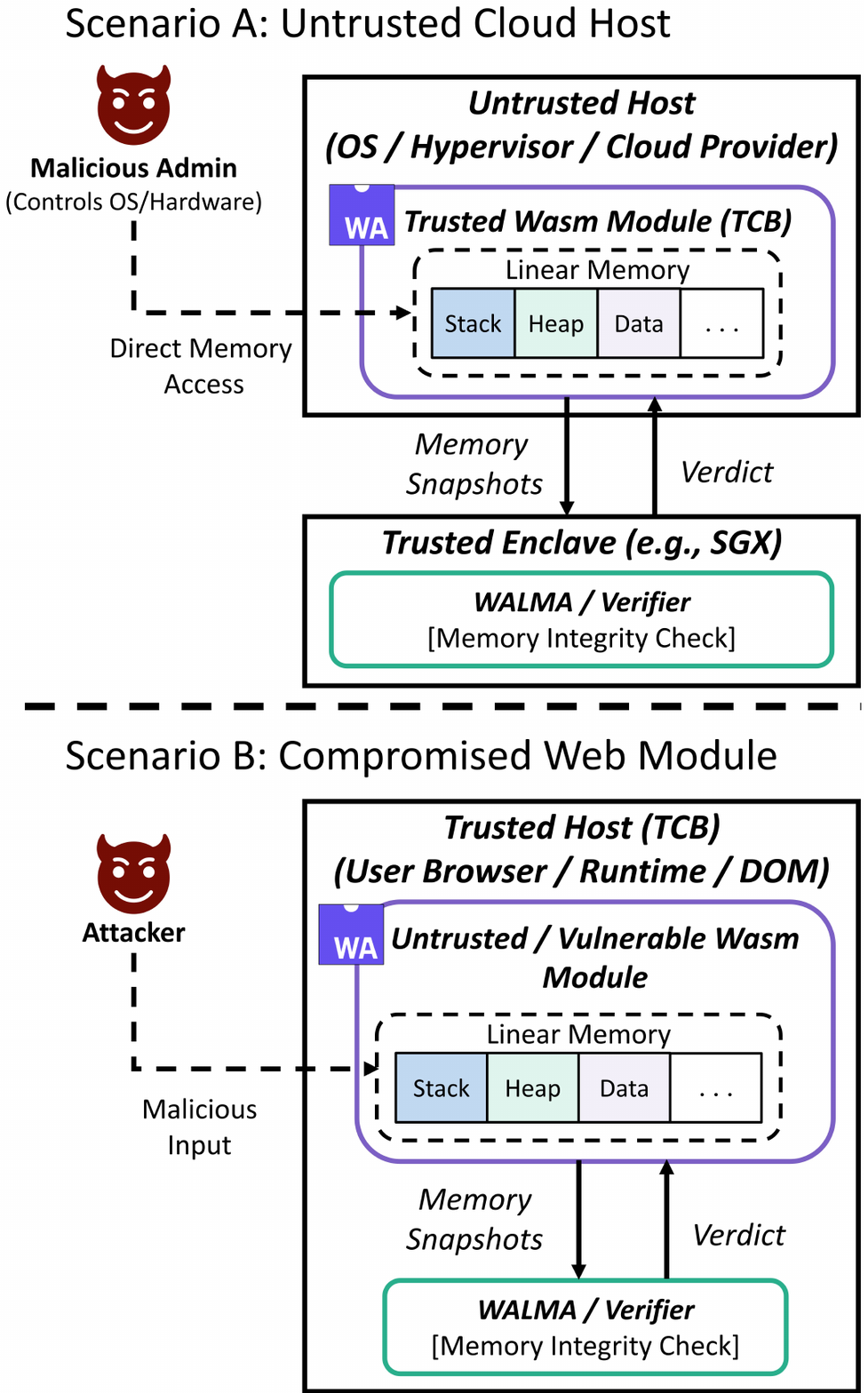}
    \caption{\tool's two deployment scenarios. \emph{Left} (Scenario A): trusted module, untrusted host; an enclave-resident verifier detects host tampering with module memory. \emph{Right} (Scenario B): trusted browser host, untrusted module; a browser-side verifier detects corruption before it reaches host state.}
    \label{fig:deployment}
\end{figure}

Wasm's trust relationship is bidirectional: a module may be the victim of its environment, or a threat to it.
We consider two deployment scenarios (\Cref{fig:deployment}) capturing both directions, each reflecting established adversarial models in cloud~\cite{tee1,tee2} and web~\cite{wemby,everything,chasms} security research.
Existing defenses mainly address one direction at a time; our key innovation is a single attester--verifier protocol covering both, with the verifier inside the trusted domain.

\subsubsection*{Scenario A: Untrusted Cloud Host}
A user deploys a trusted Wasm workload to third-party cloud or edge infrastructure (\Cref{fig:deployment}, left), where the untrusted \emph{Host} (OS, hypervisor, and physical administrators~\cite{tee1,tee2}) holds full hardware control.
Rather than escaping the sandbox through Wasm-specific data-only attacks~\cite{wemby,everything,chasms}, it attacks \emph{into} the sandbox, manipulating the workload's internal state (e.g.,\ bypassing an access-control check).
It uses system-level privileges or hardware-level tampering such as DMA~\cite{thunderclap}.
\tool measures snapshots of linear memory at instrumented execution points; only the verifier and its verdict are enclave-resident, while the Wasm runtime and the memory it manages stay on the untrusted host.
This placement is deliberate: enclave-resident memory would leave nothing to detect, while on the host it stays exposed to precisely the tampering the verifier surfaces.
Two assumptions bound this adversary.
First, \emph{the measurement path is trusted}: the host tampers with memory at will but not with snapshot capture and delivery.
This is the root-of-trust assumption of software-based attestation~\cite{swatt,pioneer}, and verifier-controlled DMA capture~\cite{dmanplay} can realize it.
It is an assumption, not a result: our prototype captures snapshots through the runtime's embedding API and does \emph{not} implement a DMA-based capture path.
For Scenario~A, \Cref{sec:eval} therefore measures the classifier \emph{given} an intact measurement path, and quantifies nothing about how to build one.
Confidential computing is complementary: a TEE or confidential VM protects the workload's memory from the platform, but says nothing about whether the state \emph{inside} it is corrupted, e.g., by an in-module bug the enclave faithfully executes.
\tool attests that state, and applies unchanged inside a confidential VM.
Second, \tool attests state at capture instants, and corruption reverted between captures is therefore a freshness gap that \Cref{sec:discussion} analyzes.
Training runs offline on developer hardware and ships the model as part of the deployment image.

\subsubsection*{Scenario B: Compromised Web Module}
In the browser, the trust relationship is inverted~\cite{wemby,everything,chasms}: the browser forms the trusted computing base, while the Wasm module is potentially vulnerable or already compromised by attacker-controlled inputs (\Cref{fig:deployment}, right).
The adversary exploits a memory corruption bug inside the module and uses the corrupted state to escalate against the embedding application.
The host trusts module data and uses it directly, e.g.,\ passing a Wasm-side string to \verb|eval()| or into the DOM via \verb|innerHTML|, a pattern found on thousands of websites~\cite{wemby}; an overflow that rewrites the string thus yields code execution in the embedding page~\cite{everything,wemby}.
Here, \tool runs as a browser-side monitor, inspecting the module's memory at instrumented points and detecting corruption inside the sandbox before it crosses into shared host state.
Unlike Scenario~A, this scenario needs no assumption on the measurement path: the trusted browser runtime captures the snapshots itself, which is the path our prototype implements.
As in Scenario A, the publisher trains the per-module model offline and ships it with the module.

\subsubsection*{Requirements}
Together these scenarios impose four requirements on any viable defense.
The defender controls neither the browser engine nor the cloud host. The defense must therefore be \circled{1}~\emph{runtime-agnostic} and add no custom Wasm runtime.
The trust relationship runs both ways. A single mechanism must therefore provide \circled{2}~\emph{bidirectional coverage}, placing the verifier inside whichever domain is trusted.
In both scenarios the adversary corrupts data while leaving the control-flow graph and the binary intact. The defense must therefore \circled{3}~\emph{detect data-centric attacks} that code- and control-flow attestation miss.
We target corruption that changes a region of memory; a targeted edit of a single flag or byte is out of scope (\Cref{sec:discussion}).
Finally, a browser verdict is latency-critical whereas a cloud deployment requires a short reaction window. The defense must therefore \circled{4}~expose a \emph{tunable resilience--performance contract}.

\section{Design}
\label{sec:design}

\begin{figure*}[tb]
    \centering
    \includegraphics[width=.8\textwidth,trim={0 4cm 0 4cm},clip]{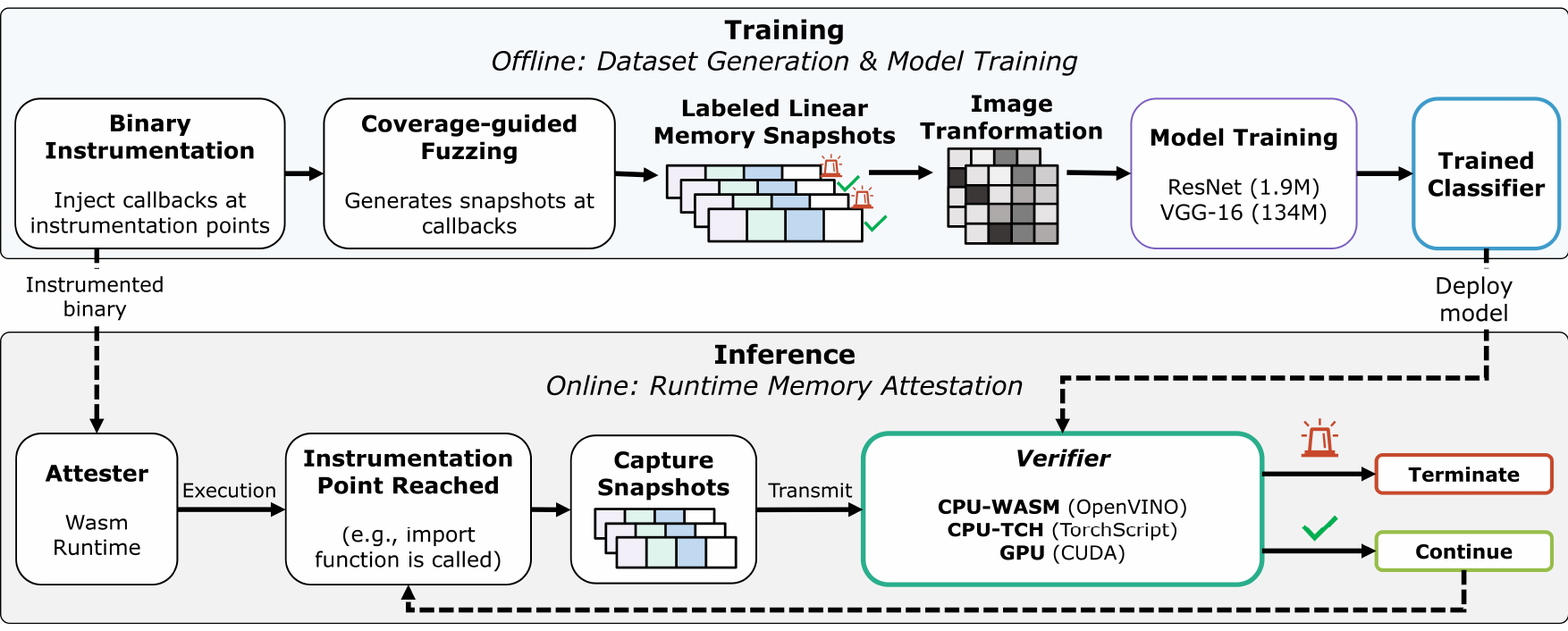}
    \caption{The \tool pipeline. \emph{Top}: offline training fuzzes each target under a sanitizer oracle to label memory snapshots and train the classifier. \emph{Bottom}: online inference streams snapshots from the deployed module to the verifier, which returns an integrity verdict.}
    \label{fig:pipeline}
\end{figure*}

\tool meets the four requirements of \Cref{sec:threat_model} by lifting integrity verification out of the binary into a learned classifier over the module's memory state.
Our architecture spans two phases (\Cref{fig:pipeline}).
An offline \emph{training} phase constructs the classifier from synthetically generated states.
In the online \emph{inference} phase, an \emph{attester} co-located with execution streams snapshots to a \emph{verifier} issuing integrity verdicts.

\subsection{Offline Phase: Classifier Construction}
\label{sec:design-offline}

The offline phase builds the classifier the runtime verifier consults, once per target on developer-controlled hardware, and ships its weights in the deployment image.
Training is therefore outside the runtime trust boundary in \emph{both} scenarios; only the inference path needs runtime protection.

\subsubsection*{Dataset Generation}
Valid Wasm memory layouts are application-specific, and therefore no closed-form rule separates a benign state from a corrupted one. \tool must learn that boundary from observed executions.
The corpus comes from binary instrumentation that injects snapshot hooks at deterministic points, plus coverage-guided fuzzing~\cite{libafl,kim2019hydra,wemby,aflpp} that drives the instrumented binary along diverse paths.
Deterministic hooks ensure that training-time snapshots follow the distribution the runtime verifier will later see.
We fuzz rather than replay the project's test suite: a classifier needs corrupted states in volume and variety, and test inputs exercise only benign paths.
We label each input as \emph{corrupted} or \emph{benign} with AddressSanitizer (ASan)~\cite{asan}, the standard oracle for memory corruption detection~\cite{libafl,aflpp}, whose near-zero false-positive rate keeps label noise out of the corrupted class.
For targets without source, the Wasm-native WBSan~\cite{wbsan} substitutes at a higher false-negative rate.
Each captured snapshot inherits the label of its input, yielding a per-target corpus of labeled snapshots.
The sanitizer labels only the corrupted class; the benign class alone defines valid memory, a fact whose effect \Cref{sec:eval-detection} quantifies.

\subsubsection*{Program-Specific Models}
\tool trains one model per program by design: the model generalizes to unseen inputs of the same program, the regime \Cref{sec:eval-detection} measures, but not across programs.
What valid memory looks like depends on a program's data structures and allocation pattern, so each model learns the layout of one program, much as control-flow attestation needs the control-flow graph of each binary~\cite{cflat}.
A leave-one-out probe supports this choice: a model trained on four targets and tested on the fifth degrades sharply, down to chance on some targets.
Its accuracy is \num{75.8}\,\% on \texttt{pal2rgb} but \num{49.7}\,\% on \texttt{pnm2png}, and its benign recall drops to \num{12}\,\% on \texttt{pdfresurrect}.
The price is an offline cost per program, amortized over deployment and reported in \Cref{sec:eval-setup}; a layout-changing rebuild needs retraining.

\subsubsection*{Image Transformation}
\begin{figure}[htbp]
    \centering
    \footnotesize
    \setlength{\tabcolsep}{1.5pt}
    \begin{tabular}{@{}cccc@{}}
        \includegraphics[width=.2\columnwidth]{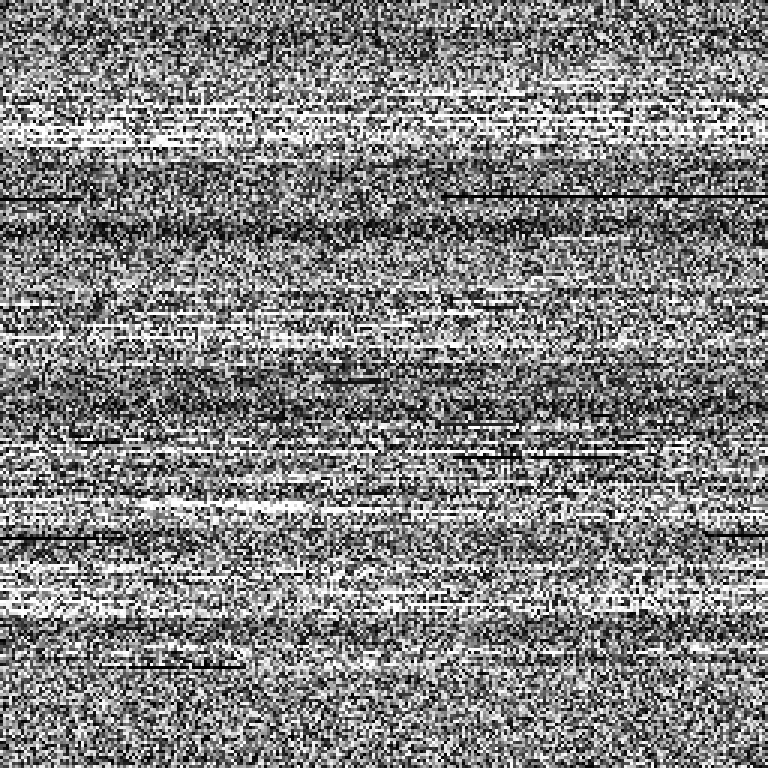} &
        \includegraphics[width=.2\columnwidth]{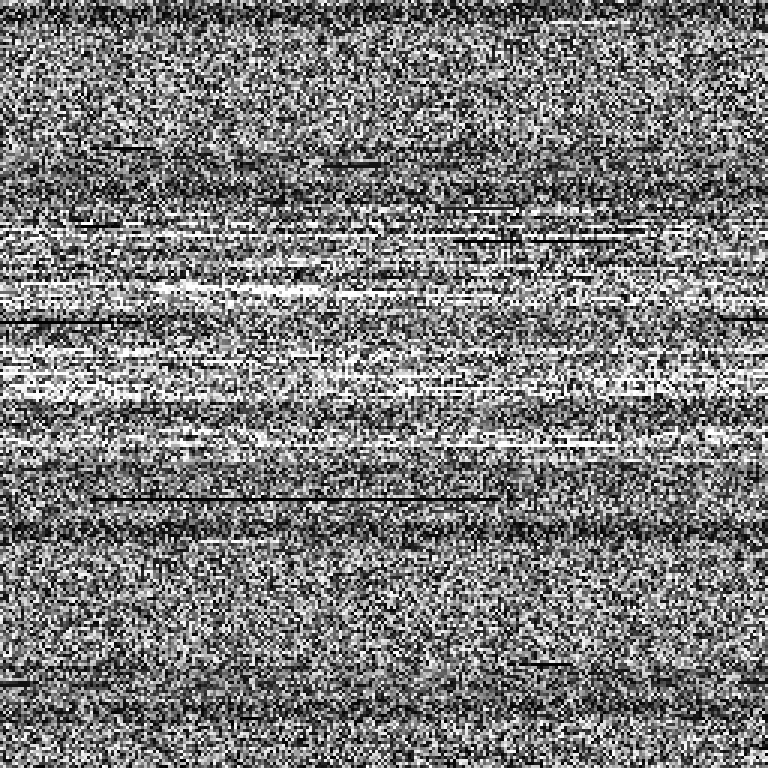} &
        \includegraphics[width=.2\columnwidth]{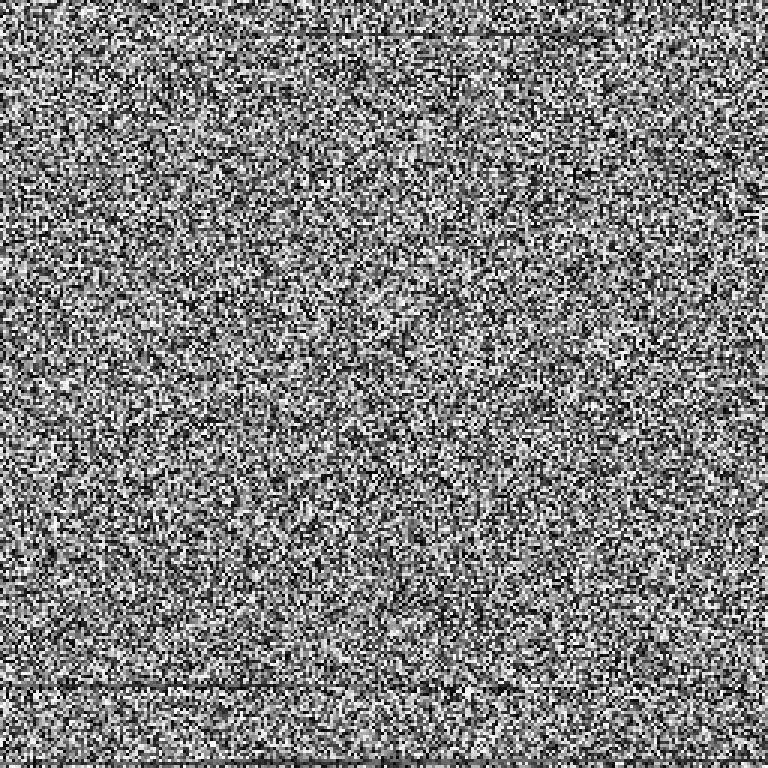} &
        \includegraphics[width=.2\columnwidth]{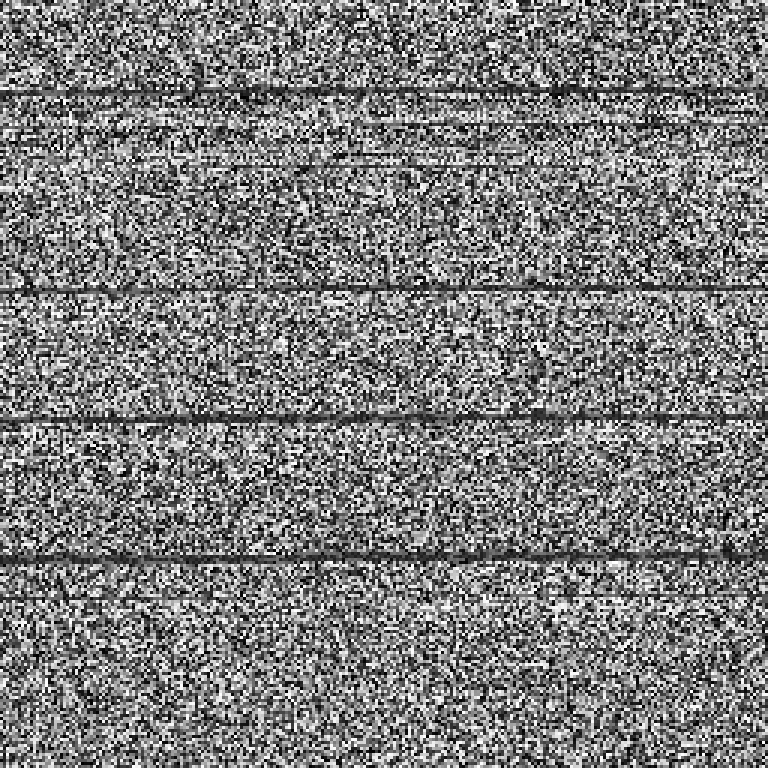} \\[1pt]
        benign & corrupted & benign & corrupted \\
        \multicolumn{2}{c}{\texttt{jbig2dec}} &
        \multicolumn{2}{c}{\texttt{pdfresurrect}} \\
    \end{tabular}
    \caption{Linear memory as a width-\num{256} grayscale image for benign and corrupted runs of two targets. Corruption perturbs the texture the CNN reads: the shift is sharp on \texttt{jbig2dec}'s regular, banded layout, but \texttt{pdfresurrect}'s interleaved PDF objects stay irregular even when benign---the hardest target in \Cref{sec:eval-detection}.}
    \label{fig:mem_images}
\end{figure}

\tool interprets the captured linear memory as a grayscale image: byte values become pixel intensities and the byte stream is reshaped at a fixed row width of \num{256} pixels.
This byteplot encoding is well-established for classifying static binaries~\cite{nataraj2011malware}; we adapt it to live memory and fix the row width from domain knowledge, making the detector knowledge-informed rather than purely data-driven.
A power-of-two width matches allocator size classes and compiler alignment rules. Repeated record-style structures---audio frames, pixel rows, allocator headers---therefore recur at fixed column offsets and form coherent stripes that a CNN receptive field can detect.
The same width also lets us reuse standard pretrained vision models~\cite{kim2019makesomenoise}.
We fixed the width on these grounds rather than by sweeping it, and note the trade-off it encodes: narrow widths such as \num{8} or \num{16} bytes place the same record on many short rows and destroy the two-dimensional locality a convolutional receptive field is built to read, collapsing the encoding toward the byte-statistics baselines of \Cref{sec:eval-detection}; much wider rows dilute a localized disruption across a long row.
\Cref{fig:mem_images} shows this encoding for two targets: a structured one, whose benign memory has a regular texture that corruption disrupts, and an irregular one, whose memory looks disordered even when benign.

\subsubsection*{Model Choice}
\tool ships two classifiers that span a latency--accuracy range; a deployment picks the point its attestation frequency allows.
We build on VGG-16~\cite{vgg} and ResNet~\cite{resnet}, proven backbones for grayscale byteplot classification in security~\cite{imcfn,rezende2017resnet}.
Our custom \num{1.9}M-parameter ResNet ingests the byte-to-pixel image directly; at \num{7.5}\,MB of weights and \num{5.3}\,ms per inference, it makes reaction at every host call or function entry affordable.
The higher-capacity \num{134}M-parameter VGG-16 is our main detector for accuracy.
Its half-gigabyte of weights fits the multi-gigabyte enclave page cache of current Intel Xeon Scalable platforms~\cite{sgxv2bench}, and it therefore remains enclave-deployable.
It suits Scenario B's off-critical-path host verifier, and fits Scenario A wherever the attestation frequency leaves room for its heavier inference.

\subsection{Online Phase: Runtime Attestation}
\label{sec:design-online}
\tool's runtime architecture (\Cref{fig:pipeline}, bottom) consists of an attester co-located with execution and a verifier whose placement matches the threat model.

\subsubsection*{Trust Assumptions}
\tool trusts the verifier and the measurement path of \Cref{sec:threat_model} that captures and delivers snapshots.
The trusted verifier state comprises the model weights, each snapshot once received, and the returned verdict---all inside the TEE in Scenario A and the browser in Scenario B.
The Wasm runtime and the linear memory it manages stay untrusted: the host can corrupt memory at any time, but cannot rewrite a snapshot in flight to make corrupted memory look clean.
The remaining attack surface is editing memory itself: \Cref{sec:eval-detection} shows that masking corrupted state from the classifier requires a contiguous overwrite of hundreds of kilobytes to megabytes.
\Cref{sec:discussion} addresses edits timed between captures.

\subsubsection*{The Attester}
Our attester runs on the untrusted host, co-located with execution.
It executes the instrumented binary; at each instrumentation point it reads the module's linear memory and strips the long runs of zero bytes typical of sparse Wasm heaps.
This keeps the snapshot small, easing the bandwidth bottleneck of remote attestation~\cite{sok_attestation}.

\subsubsection*{Instrumentation Policies}
\tool operationalizes resilience by reaction through three policies trading attestation frequency against inter-snapshot coverage:
\textbf{Import Function} captures a snapshot before every host call, reactively flagging corrupted module data before the host consumes it---sufficient to mitigate the bulk of reported Wasm-to-host escalation~\cite{chasms,everything,wemby}.
\textbf{Local Function} captures snapshots at Wasm function entries as a coarse approximation of control-flow monitoring.
\textbf{Memory Instruction} captures a snapshot before every memory-modifying instruction, providing fine-grained per-write attestation.
\Cref{sec:eval} quantifies this trade-off across the three policies (abbreviated \emph{Import}, \emph{Local}, and \emph{Memory}).

\subsubsection*{Verifier Backends}
\textbf{In-runtime (\texttt{cpu-wasm})} runs OpenVINO via WASI-NN~\cite{wasinn} in a dedicated verification runtime (e.g., Wasmtime). Inference therefore executes as sandboxed Wasm, and the snapshot never leaves the verification runtime's sandbox.
In Scenario A, the verification runtime lives in an enclave, minimizing the TCB exposed under SGX or comparable TEEs.
In Scenario B, the module's publisher ships the verifier as a second Wasm module beside the application.
Wasm's per-module memory isolation keeps the vulnerable module from reaching the verifier's weights, snapshot, or verdict.
\textbf{Native host (\texttt{cpu-tch})} runs LibTorch natively, off the Wasm critical path; in Scenario B this presumes the browser engine itself integrates \tool rather than the publisher shipping it with the module.
\textbf{Accelerated (\texttt{gpu})} offloads inference to a CUDA device via TorchScript; because raw memory must traverse the PCIe bus, this backend is appropriate only when the GPU lies inside the trust boundary~\cite{vasiliadis2014pixelvault}.

A corrupted verdict is reported the moment it occurs; detection adds no waiting interval an adversary could exploit between a corruption and its report.

\section{Implementation}
\label{sec:implementation}

We implemented \tool as an end-to-end pipeline of \num{2017} lines of Rust code (attester, instrumentation, runtime integration) and \num{409} lines of Python code (model definitions and inference service).
The same instrumented binary, attester, and runtime build serve every instrumentation policy and verifier backend, selected by a compile-time switch.

\subsubsection*{Instrumentation}
We instrument Wasm binaries with the Walrus library~\cite{walrus}, which performs source-free bytecode rewriting~\cite{vanput2005diablo,retrowrite} against any standards-compliant module.
The instrumented binary calls a single host-side function, \texttt{create\_memory\_snapshot}, at the policy-defined points; this one hook serves all three policies, and the rewriter chooses where to insert it.
Wasmtime's fuel mechanism caps each execution's instruction count, and a fuzzer-surfaced crash is therefore attributable to a memory violation rather than to runaway execution.

\subsubsection*{Fuzzing and Dataset Generation}
We integrate \texttt{LibAFL}~\cite{libafl} with the Wasmtime-based runtime to drive coverage-guided fuzzing.
We seed each campaign with the project's regression-test inputs and run it until the corpus reaches the per-target composition described in \Cref{sec:eval-setup}, yielding one model tuned to each application's memory layout.

\subsubsection*{Snapshot Pipeline}
The attester reads linear memory through Wasmtime's embedding API and strips null-byte runs.
When persisting snapshots for training we additionally apply chunked run-length encoding (RLE); on the online attestation path RLE is skipped to minimize latency.
The verifier reconstructs the row-aligned grayscale image by padding and reshaping the decoded byte stream to the fixed \num{256}-pixel width.
The ResNet ingests this image at full resolution, while VGG-16 resizes it to its fixed input dimensions.

\subsubsection*{Verifier Backends}
The three backends share the same length-prefixed wire format and the same model artifact.
\texttt{cpu-wasm} runs OpenVINO through the WASI-NN \texttt{load}/\texttt{init\_execution\_context}/\texttt{compute} sequence inside the verification runtime, keeping snapshot data inside the sandbox.
For Scenario A we package this runtime inside Gramine~\cite{gramine} as a separate SGX enclave, alongside the untrusted application runtime.
\texttt{cpu-tch} runs LibTorch natively on the host.
\texttt{gpu} reuses the LibTorch front-end but offloads inference to a CUDA device via TorchScript.
\section{Evaluation}
\label{sec:eval}

Our evaluation answers two questions: whether visual classification of linear memory reliably detects corruption on unseen inputs (\Cref{sec:eval-detection}), and what continuous attestation costs across standard Wasm workloads (\Cref{sec:eval-performance}).
Our prototype captures snapshots through the trusted runtime, so the detection results hold end to end for Scenario~B; for Scenario~A, they evaluate the classifier under the trusted measurement path of \Cref{sec:threat_model}.

\subsection{Experimental Setup}
\label{sec:eval-setup}

All experiments ran on an Intel Xeon w7-3565X processor with an Nvidia RTX 5000 Ada Generation GPU.
We compared the three backends described in \Cref{sec:design-online}, running each configuration \num{25} times per binary.
As in Scenario A, \texttt{cpu-wasm} runs and is measured inside a Gramine~\cite{gramine} SGX enclave; \texttt{cpu-tch} and \texttt{gpu} run natively on the host.

We validate detection on a \emph{security corpus} of five real-world Wasm applications with known memory-safety vulnerabilities (CVEs), spanning four input domains rather than one: lossless audio coding (\texttt{flac}), palettized raster conversion (\texttt{pal2rgb}, from LibTIFF), netpbm-to-PNG conversion (\texttt{pnm2png}, from libpng), bilevel image coding (\texttt{jbig2dec}), and PDF document parsing (\texttt{pdfresurrect}).
The bug classes also differ, and they determine how corruption is laid down in memory.
\texttt{flac}, \texttt{pal2rgb}, and \texttt{jbig2dec} carry \emph{heap} buffer overflows: an undersized allocation---in \texttt{jbig2dec} and \texttt{pal2rgb} the product of an integer overflow in a size computation---is then written past its end. Corruption therefore lands in the heap and damages neighboring allocations and allocator metadata.
\texttt{pnm2png} and \texttt{pdfresurrect} carry \emph{stack} buffer overflows in header and trailer parsing, where an unbounded copy into a fixed-size local buffer overruns the shadow stack region of linear memory.
Both classes are spatial out-of-bounds writes, which is what makes them visible to a data-centric monitor at all: each leaves a contiguous, misplaced write in memory that persists after the faulting instruction retires.
We return in \Cref{sec:discussion} to the classes this excludes.
These bugs are fixed upstream, and we use the affected releases deliberately: a known, sanitizer-confirmed vulnerability supplies ground-truth labels at volume.
Patching them would remove the corpus, not the problem \tool addresses, which is the bug that is still unpatched or still unknown in a module already deployed.
All five are batch-style parsers and codecs that terminate, which suits fuzzing.
We measure runtime overhead on WABench~\cite{wabench}, \num{53} binaries in total: \num{30} PolyBench~\cite{polybench} compute kernels, \num{12} MiBench~\cite{mibench} embedded utilities, and \num{11} real-world applications with large memory needs.

\subsubsection*{Corpus Construction and Split}
For each target, we ran a \texttt{LibAFL}~\cite{libafl} fuzzing campaign on a binary instrumented with ASan~\cite{asan}, seeded with the project's standard test inputs.
We labeled an input \emph{corrupted} if it raised an ASan error and \emph{benign} if it completed cleanly.
ASan serves only to label inputs: we capture the snapshots by replaying each input on the uninstrumented binary, where nothing intercepts the out-of-bounds write and it lands in linear memory.
Wasm does not trap on writes inside linear memory, so execution continues on the corrupted state, and the model classifies that state rather than a crash.
We ran the fuzzer until we collected \num{1000} corrupted and \num{1000} benign snapshots per target; caps on execution time and snapshot frequency prevent long benign runs from flooding the corpus with duplicates.
Crucially, we split training and testing data by program \emph{input}, not by snapshot.
The test set holds \num{50} ASan-flagged and \num{50} benign inputs the model never saw during training.
The model therefore cannot memorize a specific input; it must detect corruption on unseen inputs of the same program.
Labels are input-granular: every snapshot of a corrupted run inherits the corrupted label, including captures taken \emph{before} the violation occurs.
This over-labels, and we keep it deliberately: it favors corpus volume over label precision, and it biases the reported numbers against us, since a pre-violation snapshot of a corrupted run is indistinguishable from a benign one and can only ever be scored as a miss.
Selecting only post-violation snapshots would raise the reported accuracy without changing what the model detects; the ablations below show verdicts key on memory state, not inputs, either way.

\begin{table*}[t]
    \centering
    \small
    \setlength{\tabcolsep}{10pt}
    \caption{Per-target detection (\%) over per-snapshot verdicts at the \emph{Import} policy. \emph{Hist.} (byte histogram), \emph{Size} (dump size), and \emph{Texture} (texture descriptors) are non-CNN baselines; F1 is benign-class. \emph{FPR}: benign-run snapshots flagged corrupted; \emph{FNR}: corrupted-run snapshots passed benign (attack recall $=1-$FNR).}
    \label{tab:accuracy}
    \begin{tabular}{lrrrrrrrrrrr}
        \toprule
        & \multicolumn{3}{c}{\shortstack{\textbf{Baseline F1}\\[1pt]{\scriptsize non-CNN}}}
        & \multicolumn{4}{c}{\shortstack{\textbf{ResNet}\\[1pt]{\scriptsize \num{1.9}M, in-runtime}}}
        & \multicolumn{4}{c}{\shortstack{\textbf{VGG-16}\\[1pt]{\scriptsize \num{134}M, main}}} \\
        \cmidrule(lr){2-4} \cmidrule(lr){5-8} \cmidrule(lr){9-12}
        \textbf{Target} & Hist. & Size & Texture & Acc. & F1 & FPR & FNR & Acc. & F1 & FPR & FNR \\
        \midrule
        flac         & 54.4 & 97.2 & 98.6 & 100.0 & 100.0 & 0.0  & 0.0  & 100.0 & 100.0 & 0.0 & 0.0  \\
        pal2rgb      & 64.4 & 82.0 & 84.3 & 100.0 & 100.0 & 0.0  & 0.0  & 100.0 & 100.0 & 0.0 & 0.0  \\
        pnm2png      & 74.8 & 74.9 & 65.0 & 94.0  & 93.9  & 8.0  & 4.0  & 97.0  & 97.1  & 0.0 & 6.0  \\
        jbig2dec     & 72.8 & 63.1 & 85.3 & 94.0  & 94.3  & 0.0  & 12.0 & 95.0  & 95.2  & 0.0 & 10.0 \\
        pdfresurrect & 49.5 & 43.6 & 69.5 & 66.0  & 69.6  & 22.0 & 46.0 & 74.0  & 79.4  & 0.0 & 52.0 \\
        \bottomrule
    \end{tabular}
\end{table*}

\subsubsection*{Cost of the Offline Phase}
Per target, the offline phase comprises one fuzzing campaign and one training run on developer hardware, outside the runtime trust boundary.
Training is short: VGG-16 converges in four epochs and the ResNet in five, on a single GPU.
The resulting artifacts differ by two orders of magnitude: \num{7.5}\,MB for the ResNet against roughly \num{0.5}\,GB for VGG-16.
Per-target snapshot sizes are given in \Cref{fig:adversarial_evasion}.
We did not instrument campaign wall-clock time; the fuzzing campaign, not training, dominates the offline cost.

\subsection{Detection Effectiveness}
\label{sec:eval-detection}

\subsubsection*{Necessity of the Visual Encoding}
We test whether treating memory as an image finds corruption that byte statistics miss, against three baselines (\Cref{tab:accuracy}).
\emph{Hist.} is a normalized \num{256}-bin byte-frequency histogram, blind to dump size.
\emph{Size} uses only the dump size, which differs up to \num{43}$\times$ between benign and corrupted states on \texttt{flac}.
\emph{Texture} classifies the same width-\num{256} grayscale image our CNN uses with hand-crafted texture descriptors, isolating what deep learning adds.
It follows the classical image-based malware classification pipeline~\cite{nataraj2011malware}: local binary pattern and Haralick co-occurrence features feed a linear classifier.
We do not compare against a full prior system, because none attests the linear memory of a Wasm runtime.
The closest data-integrity attestation schemes~\cite{dialed,oneforall,dmanplay} target microcontroller firmware and protect hand-picked variables, so they cannot run on our targets without a redesign.
ASan is our labeling oracle and matches the labels by construction; \Cref{sec:discussion} compares both.

\emph{Texture} is the intermediate point between a single integer and a deep network, and the strongest baseline on almost every target. The memory image therefore carries security signal that even non-learned algorithms can extract, and the encoding---not the model class---is doing much of the work.
VGG-16 nonetheless beats all three on every target, by up to \num{22} F1 points, which is the margin that justifies the learned model over the cheaper descriptor.
It wins where \emph{Size} is weak (\texttt{jbig2dec}, \texttt{pnm2png}, \texttt{pdfresurrect}) because it reads local layout rather than dump size, and adds the decisive margin even where size alone is informative (\texttt{flac}, \texttt{pal2rgb}).

\subsubsection*{Per-Target Detection}
We measure detection at the \emph{Import} operating point.
Each test run captures a snapshot at every host call, and every resulting verdict is scored against the ground-truth label of the input that produced the run.
The per-snapshot metrics in \Cref{tab:accuracy} show that detection tracks how cleanly a program lays out its data.
Well-structured targets like \texttt{flac} and \texttt{pal2rgb}, whose corruption visibly breaks audio frames and pixel rows, reach F1 above \num{99} with zero false positives and zero missed attacks under both models.
The document parser \texttt{pdfresurrect} is the hard case: PDFs interleave variable-sized objects, and its memory therefore looks irregular even when benign.
The in-runtime ResNet holds the measured false-positive rate (FPR) at zero on \texttt{flac}, \texttt{pal2rgb}, and \texttt{jbig2dec}, rising to \num{8}\,\% on \texttt{pnm2png} and \num{22}\,\% on \texttt{pdfresurrect}.
The larger VGG-16 flags no benign snapshot on any target.
Because per-snapshot false positives compound over a run, this zero FPR means no benign test input raises an alarm at all---the property that lets an administrator act on alerts automatically.
Its cost is a higher miss rate on the hard target (false-negative rate \num{52.0}\,\% vs.\ ResNet's \num{46.0}\,\% on \texttt{pdfresurrect}).
Where the in-runtime ResNet's FPR is nonzero, automated response instead calls for verdict aggregation, e.g., alarming on $k$ consecutive corrupted verdicts.
With \num{100} test inputs per target the confidence intervals are wide, but the ranking holds across targets.

\subsubsection*{Externally Injected Corruption}
A malicious administrator in Scenario A can write directly into linear memory out-of-band (e.g., via DMA~\cite{thunderclap,dmanplay}) with no triggering input.
We simulate this by scattering random bit-flips into benign snapshots, scaled from roughly \num{512}\,B to \num{107}\,KB.
To isolate what lets \tool recognize this out-of-distribution tampering, we ablate the VGG-16 verifier's training data.
We compare an \emph{ASan-only} model, trained solely on the fuzzer-derived corpus, against a \emph{+Injection} model whose corrupted class is augmented with \num{500} such injected snapshots.

\begin{table*}[htbp]
    \centering
    \small
    \setlength{\tabcolsep}{10pt}
    \caption{Learnability of externally injected corruption as held-out balanced accuracy; \emph{Entropy} is benign-memory Shannon entropy in bits per byte.}
    \label{tab:scenario_a_learnability}
    \begin{tabular}{lrrrrrrrrrrr}
        \toprule
        & & \multicolumn{5}{c}{\textbf{ASan-only}} & \multicolumn{5}{c}{\textbf{+Injection}} \\
        \cmidrule(lr){3-7} \cmidrule(lr){8-12}
        \textbf{Target} & \textbf{Entropy} & 1\% & 5\% & 10\% & 25\% & 50\% & 1\% & 5\% & 10\% & 25\% & 50\% \\
        \midrule
        pnm2png      & 2.65 & 0.65 & 0.80 & 0.85 & 0.95 & 1.00 & 0.95 & 0.95 & 0.90 & 0.95 & 0.95 \\
        pal2rgb      & 6.96 & 0.60 & 0.55 & 0.55 & 0.65 & 0.90 & 0.60 & 0.60 & 0.60 & 0.55 & 0.60 \\
        jbig2dec     & 7.77 & 0.65 & 0.65 & 0.65 & 0.70 & 0.65 & 0.65 & 0.70 & 0.75 & 0.65 & 0.65 \\
        pdfresurrect & 7.93 & 0.60 & 0.55 & 0.50 & 0.60 & 0.50 & 0.50 & 0.50 & 0.50 & 0.50 & 0.50 \\
        flac         & 7.94 & 0.50 & 0.50 & 0.50 & 0.50 & 0.50 & 0.50 & 0.55 & 0.50 & 0.60 & 0.50 \\
        \bottomrule
    \end{tabular}
\end{table*}

\Cref{tab:scenario_a_learnability} reports detectability as held-out balanced accuracy, and the result is entropy-bound.
The ASan-only model already flags injected corruption on the low-entropy target \texttt{pnm2png} (\num{0.85} once a tenth of the bits flip, \num{1.00} at half).
This is a cross-generator result: the model never saw injected corruption during training.
\emph{+Injection} sharpens this to \num{0.95} from just \num{1}\,\% corruption.
On high-entropy targets near \num{8} bits per byte (\texttt{flac}, \texttt{pdfresurrect}), random flips barely shift an already-random distribution, and global average pooling absorbs isolated changes, leaving both models near chance.
This entropy bound applies to out-of-band injection, not to the exploit-induced corruption of \Cref{tab:accuracy}: \texttt{flac}, at \num{7.94} bits per byte, is detected there without error, because a real overflow changes the layout and size of memory, not only its byte values.
A deployer can measure benign-memory entropy from benign runs alone, before deployment, to judge whether this out-of-band coverage applies to a target.
Flips at Rowhammer granularity~\cite{jattke2024zenhammer}---single bits rather than percent-scale budgets---fall below the smallest tested budget and below the detection floor on every target.

\subsubsection*{Model Evasion via Targeted Perturbations}
\begin{figure}[htbp]
    \centering
    \includegraphics[width=.85\columnwidth]{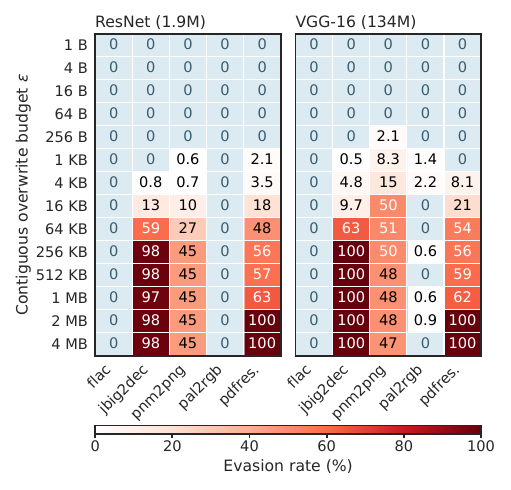}
    \caption{Black-box contiguous-overwrite evasion for the in-runtime ResNet (\emph{left}) and main VGG-16 (\emph{right}); shade encodes evasion rate, pale-blue (0) no successful evasion. Median snapshot size, which caps $\varepsilon$: \texttt{flac} \num{1.5}\,MB, \texttt{jbig2dec} \num{71}\,KB, \texttt{pnm2png} \num{39}\,KB, \texttt{pal2rgb} \num{9}\,KB, \texttt{pdfresurrect} \num{346}\,KB.}
    \label{fig:adversarial_evasion}
\end{figure}

An attacker who edits memory before a capture might perturb the snapshot to hide corruption already present, in the spirit of adversarial examples~\cite{fgsm,capozzi2025robustness}.
The opportunity is scenario-dependent: Scenario B captures memory in the browser beyond the attacker's reach, and a Scenario A edit below the entropy floor of \Cref{tab:scenario_a_learnability} needs no evasion to begin with.
We therefore test the complementary case---hiding corruption large enough to be seen---on all targets with both models.
Starting from a corrupted snapshot, the attacker flips bytes up to a budget $\varepsilon$ to induce a benign verdict.
Scattered flips rarely succeeded.
A contiguous block overwrite (\Cref{fig:adversarial_evasion}), swept from \num{1}\,byte to \num{4}\,MB (about \num{20}\,\% of memory), worked only at large sizes.
\texttt{flac} and \texttt{pal2rgb} never exceeded a \num{2.5}\,\% evasion rate even at \num{4}\,MB; the remaining targets reached full evasion only after contiguous overwrites of hundreds of kilobytes to megabytes---a large fraction of their total memory---with partial evasion appearing from \num{16}--\num{64}\,KB.
Targeted perturbation therefore fails: evasion is target-dependent and demands massive overwrites that are easy to spot.

\subsubsection*{The Sanitizer Only Labels the Training Data}
Two of the preceding results share one cause.
A snapshot overwritten entirely with random bytes carries no ASan label and matches no corrupted training run, yet \tool rejects it on every target: it looks nothing like the valid-memory prior that the benign class defines.
The injected corruption of \Cref{tab:scenario_a_learnability} is the same effect under a subtler perturbation---no triggering input, no sanitizer label---and is still caught on the structured target \texttt{pnm2png}.
The sanitizer is therefore a floor on \tool's coverage, not a ceiling: it labels the training data, but the verdict measures departure from valid memory rather than the presence of a sanitizer-visible fault.
\Cref{tab:scenario_a_learnability} also bounds where this learned prior is sharp enough to use: below roughly \num{3}\,bits per byte of benign entropy it separates tampered from valid state.
Near the \num{8}-bit ceiling (\texttt{flac}, \texttt{pdfresurrect}), benign and tampered memory are statistically indistinguishable under the bytes-as-pixels encoding; these targets need the locality-preserving encodings of \Cref{sec:discussion}, not a new attestation mechanism.

\subsection{Performance}
\label{sec:eval-performance}
\begin{figure*}[t]
    \centering
    \includegraphics[width=.9\textwidth]{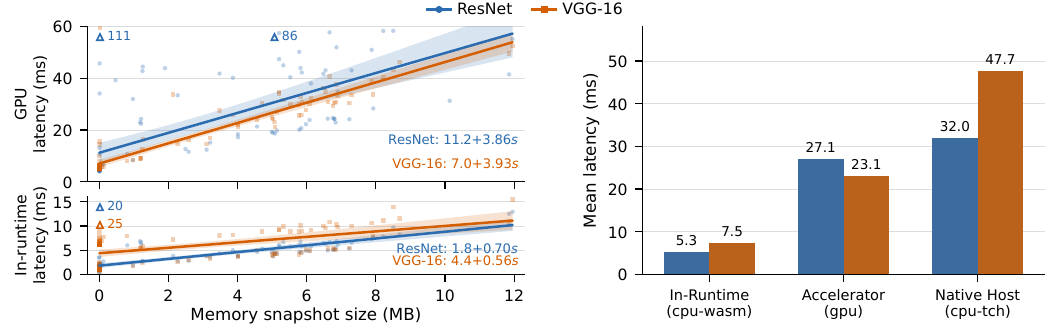}
    \caption{\emph{Left}: off the runtime, latency rises five to seven times faster per megabyte, and in-runtime inference therefore wins at every snapshot size by a widening margin. Points are individual runs, bands \num{95}\,\% CIs, four off-scale runs labelled at top. \emph{Right}: \texttt{cpu-wasm} up to $6\times$ faster than off-runtime backends.}
    \label{fig:inference_performance}
\end{figure*}

\subsubsection*{Inference Latency and Memory Scaling}
Inference latency grows with snapshot size and differs sharply across backends (\Cref{fig:inference_performance}).
The in-runtime \texttt{cpu-wasm} backend runs a ResNet check in \num{5.3}\,ms, against \num{27.1}\,ms on the GPU and \num{32.0}\,ms on \texttt{cpu-tch}. VGG-16 costs \num{7.5}\,ms in-runtime, \num{23.1}\,ms on GPU, and \num{47.7}\,ms on \texttt{cpu-tch}.
The ordering is counterintuitive: for models this small, moving megabytes across the PCIe bus costs more than the inference---\num{3.9}\,ms per megabyte off the runtime against \num{0.6}\,ms in it---and the local CPU backend is therefore the fastest of the three.
The performance-optimal choice is also the one that never exposes snapshots off the runtime.

\subsubsection*{Overhead Across Workload Classes}
Averaged over \num{25} runs of each of the \num{53} binaries (\Cref{tab:walma_overhead_classes}), the \emph{Import} policy---checking memory exactly where an attack escapes the sandbox---incurs only a \num{1.07}$\times$ slowdown on compute kernels and \num{1.69}$\times$ on host-call-heavy workloads.
\emph{Local} checks every function call (\num{1.50}$\times$/\num{2.05}$\times$/\num{2.65}$\times$ across the three suites) and \emph{Memory} nearly every write (\num{1.78}$\times$--\num{7.57}$\times$, an upper bound).
This realizes the tunable resilience--performance contract of \Cref{sec:threat_model}: a shorter reaction window costs proportionally more.
In-runtime placement compounds this as checks grow frequent: under \emph{Memory} on full applications, \texttt{cpu-wasm} holds \num{7.22}$\times$ while off-runtime backends exceed \num{15}$\times$.
Whether \num{1.07}$\times$--\num{1.69}$\times$ is acceptable is a deployment question, which is why three policies exist.
The comparison class is not an unmonitored binary but deployed sanitizers, which cost more and, as \Cref{sec:discussion} argues, cannot cover Scenario~A in principle.

\begin{table}[htbp]
  \centering
  \setlength{\tabcolsep}{6pt}
  \caption{Runtime overhead by workload class, VGG-16 verifier; geometric mean over \num{53} binaries.}
  \label{tab:walma_overhead_classes}
  \begin{tabular}{llrrr}
    \toprule
    \textbf{Backend} & \textbf{Policy} & \textbf{PolyBench} & \textbf{MiBench} & \textbf{Full-Apps} \\
    \midrule
    \texttt{cpu-wasm} & Import  & 1.07 & 1.69 & 1.59 \\
     & Local   & 1.50 & 2.05 & 2.65 \\
     & Memory  & 1.78 & 7.57 & 7.22 \\
    \midrule
    \texttt{cpu-tch} & Import  & 1.15 & 1.59 & 1.79 \\
     & Local   & 2.71 & 1.85 & 5.34 \\
     & Memory  & 3.43 & 5.95 & 15.51 \\
    \midrule
    \texttt{gpu} & Import  & 1.13 & 1.49 & 1.88 \\
     & Local   & 2.31 & 1.97 & 4.51 \\
     & Memory  & 3.81 & 6.06 & 15.12 \\
    \bottomrule
  \end{tabular}
\end{table}

\section{Discussion and Limitations}
\label{sec:discussion}

\subsubsection*{Extending Coverage}
\tool's weak targets are those whose benign memory lacks row-aligned structure (\Cref{sec:eval-detection}).
\emph{Alternative encodings} (Hilbert-curve or Z-order linearization) expose non-row locality that bytes-as-pixels hides.
\emph{Alternative learners} without a two-dimensional locality assumption (byte-level Transformer or recurrent models) better fit text-shaped memory.
\emph{One-class and semi-supervised learners}, such as autoencoders or Deep SVDD~\cite{ruff2018deep}, trained on benign snapshots alone would remove the need for a corruption oracle during training.
Benign-memory drift in long-running services poses the same challenge over time.

\subsubsection*{Evaluation Scope}
Our corrupted class comes from fuzzer-found inputs that trigger each CVE, on only five targets with \num{100} held-out inputs each.
Such bugs enable host code execution~\cite{everything,wemby}; whether our five CVEs do is future work, as we built no exploits.
A full exploit drives the same out-of-bounds write with a different payload, and \Cref{sec:eval-detection} shows the verdict keys on memory state, not input; only a few-byte payload goes undetected, as discussed next.

\subsubsection*{Adaptive Adversaries}
A privileged host with white-box model access could mount targeted-perturbation evasion~\cite{fgsm,capozzi2025robustness}, but the Scenario A enclave denies that access, forcing the black-box attack of \Cref{sec:eval-detection}, where masking corruption demands a wholesale overwrite under both models.
The sharper limit is at the other end of the size range: a targeted data-only edit~\cite{dop2}---one flag, one length field, or the single byte an off-by-one overflow writes---is \emph{not} detected by \tool, because it falls below the \num{1}\,\% budget of \Cref{tab:scenario_a_learnability} and global average pooling averages an isolated change away before it can move a logit.
This bounds the security claim to corruption with a spatial footprint and excludes precise data-only manipulation, the regime where per-variable defenses~\cite{dmanplay,oneforall} remain the right tool and compose with ours.
Closing it calls for a change of granularity rather than a better classifier, through per-region attestation over per-allocation sub-images.
Side-channel leakage of enclave-resident models~\cite{lehocky2025openvino} could partially restore white-box knowledge, and per-tenant model diversity hardens that residual risk.

\subsubsection*{The Sanitizer Alternative}
One might object that the binary could carry AddressSanitizer, localized to fuzzer-identified sites, instead of snapshot hooks.
Three differences separate them.
First, a sanitizer executes \emph{inside} the module; in Scenario~A, a compromised host can simply run the uninstrumented binary or patch the checks out.
Second, sanitizers only intercept violating \emph{instructions}, making external writes to linear memory invisible by construction (\Cref{tab:scenario_a_learnability}).
Third, source-free sanitization (e.g., WBSan~\cite{wbsan}) trades high runtime overhead and false negatives for compiler independence, whereas \tool's \emph{Import} policy incurs \num{1.07}$\times$--\num{1.69}$\times$ without source, scaling with host exposure rather than with runtime.

\subsubsection*{Snapshot Freshness}
\tool classifies each snapshot independently, with no nonce or challenge-response protocol, and therefore does not by itself resist replay or a corrupt-then-restore race.
A Scenario A host can flip an access-control flag after one capture and restore it before the next: every inspected snapshot is clean, yet the corrupted decision has executed.
It can also replay an earlier clean snapshot to mask persistent corruption; both are freshness matters, not limits of visual classification.
Binding each capture to a fresh nonce is a standard attestation primitive~\cite{pioneer,swatt}, and the per-write Memory policy shrinks the race window to a single instruction; we leave this integration to future work.

\subsubsection*{A General Integrity Primitive}
Because the verdict measures departure from valid memory rather than a sanitizer signature (\Cref{sec:eval-detection}), the training oracle is interchangeable.
Any labeler of corrupted states, not only ASan, defines a new policy over the same pipeline.
Where classical schemes certify the binary~\cite{swatt,pioneer} or the execution path~\cite{cflat}, \tool certifies the data state, and that state's entropy, not the oracle, bounds what it can certify.

\subsubsection*{Scope}
\tool is a detection primitive: attestation establishes whether the module's state is intact and hands the reaction to a downstream mechanism.
Prevention under Scenario A would mean trusting the host we do not trust; under Scenario B it would mean hardening the module itself, which complementary work~\cite{wbsan,fuzzm,pkuwa} addresses.
\tool composes with these defenses: its verdicts can gate a response, such as halting the module or rejecting its output, closing the loop from detection to reaction.

\section{Related Work}
\label{sec:related}

\subsubsection*{WebAssembly Memory Safety}
Compartmentalization (MSWasm~\cite{mswasm}, RLBox~\cite{rlbox}), hardware-assisted isolation (PKUWA~\cite{pkuwa}), and WASI-interface hardening (WaVe~\cite{wave}) require custom runtimes, compilers, or source annotations inapplicable to deployed modules.
Binary-rewriting sanitizers~\cite{wbsan,fuzzm} and symbolic execution~\cite{seewasm} find memory errors offline, with no recourse against an adversarial host; measurement work~\cite{chasms,everything,wemby} characterizes Wasm-to-host escalation but offers no runtime defense.

\subsubsection*{Remote and Control-Flow Attestation}
SWATT~\cite{swatt} and Pioneer~\cite{pioneer} verify static code identity; others attest device liveness~\cite{jin2019aliveness}.
Control-flow attestation (CFA)~\cite{cflat,oat,litehax,atrium} traces the executed control-flow graph but, as a recent SoK notes~\cite{sok_attestation}, misses data-only attacks~\cite{dop2} that preserve a valid control-flow path.
DIALED~\cite{dialed} and OneForAll~\cite{oneforall} attest input integrity, DMA'n'Play~\cite{dmanplay} validates selected variables over DMA, and hardware integrity trees detect tampering architecturally~\cite{vig2018skewedhashtree}, all on microcontroller-class systems with hand-picked variables.
\tool is, to our knowledge, the first to attest the full mutable state of a general-purpose runtime, replacing per-variable instrumentation with a single classifier over the memory image.

\subsubsection*{Learning over Memory and Binaries}
Rendering binaries as grayscale images is an established malware-classification technique~\cite{nataraj2011malware}, applied to memory dumps for OS-independent malware detection~\cite{Memory_CNN_Other} and to Wasm binaries for cryptojacking detection~\cite{crypto4}.
Further work recovers structure from kernel memory images~\cite{lin2011siggraph}, learns over raw binary code~\cite{massarelli2019safe}, and builds signatures from control-flow structure~\cite{bonfante2009morphological}.
All classify static artifacts; \tool instead learns reactively, over the live state of a running module rather than a binary or post-mortem dump.

\section{Conclusion and Summary}
\label{sec:conclusion}
This paper presented \tool, the first linear-memory attestation framework for WebAssembly, which reduces runtime integrity verification to image classification and covers the compromised-module direction of the Wasm trust relationship end to end, and the untrusted-host direction under a trusted measurement path.
Its VGG-16 model reaches \num{95}\,\%--\num{100}\,\% verdict accuracy on four of five CVE-affected applications without a single false alarm on benign input, at 1.07$\times$ overhead on compute-bound modules and 1.69$\times$ on host-call-heavy ones across the \num{53} WABench binaries.
\tool demonstrates that data-centric attestation delivers resilience by reaction, letting systems tailor verification granularity and runtime overhead to their threat model.

\section*{Acknowledgments}
This work was partially funded by the Deutsche Forschungsgemeinschaft (DFG,
German Research Foundation) through SFB 1119 -- 236615297, project S2.

{
    \printbibliography
}

\section*{LLM Usage Statement}
The authors used a large language model as a writing and editing aid: revising
prose for concision, converting the manuscript between document classes, and
copy-editing. It was not used to generate research ideas, to design or run any
experiment, or to produce, analyze, or interpret any measurement. Every
quantitative claim in this paper is derived from the authors' own experimental
data, and every citation was checked against the cited source by the authors,
who take full responsibility for the correctness, originality, and integrity of
the work.

\end{document}